\documentclass[pdftex,prl,amsmath,superscriptaddress,twocolumn]{revtex4}

\usepackage{graphicx}
\usepackage[utf8]{inputenc}

\begin{document}

\title{Time walkers and spatial dynamics of aging information}
\author{L. Lizana}
\affiliation{Niels Bohr Institute, Blegdamsvej 17, DK-2100, Copenhagen, Denmark}                                                                             
\author{M. Rosvall}
\affiliation{Ume\aa~University, Ume\aa, Sweden}
\affiliation{Niels Bohr Institute, Blegdamsvej 17, DK-2100, Copenhagen, Denmark}
\author{K. Sneppen}
\affiliation{Niels Bohr Institute, Blegdamsvej 17, DK-2100, Copenhagen, Denmark}                                                                             


%
%
\begin{abstract}
The distribution of information is essential for living system's ability to coordinate and adapt. Random walkers are often used to model this distribution process and, in doing so, one effectively assumes that information maintains its relevance over time. But the value of information in social and biological systems often decays and must continuously be updated. To capture the spatial dynamics of aging information, we introduce time walkers. A time walker moves like a random walker, but interacts with traces left by other walkers, some representing older information, some newer. The traces form a navigable information landscape.  We quantify the dynamical properties of time walkers moving on a two-dimensional lattice and the quality of the information landscape generated by their movements. We visualize the self-similar landscape as a river network, and show that searching in this landscape is superior to random searching and scales as the length of loop-erased random walks.
\end{abstract}

\maketitle

%
%

The adaptation and organization of living systems rely on communication; a constant distribution of information allows biological and social systems on all scales to adjust and synchronize to their surroundings. Pairwise and direct communication is simple to manage. In many systems, however, from signal transduction cascades in bacteria \cite{camilli2006bacterial} and cell-to-cell signaling during development of multicellular organisms \cite{trosko1998cell} to everyday gossiping in human societies \cite{dunbar2004gossip} and continually updated routing tables in networks \cite{katz2002adaptation}, intermediate steps take place between multiple senders and receivers. Consequently, in the resulting confusion of crosstalk, living systems must have a systematic way of evaluating information to distinguish updated from outdated.

The notion of information can be ambiguous, but many social, biological, and financial systems have in common that: 1) information gives a competitive advantage, 2) new information is more valuable than old information, and 3) information can be replicated and communicated. That is, for any strategic choice with uncertainty involved, access to information can reduce the uncertainty and be turned into power \cite{stigler1961economics}; in a world with fluctuating interests, environments, and prices, the value of information inevitably decays with time \cite{stiglitz1981credit}; and, finally, in the absence of public information, there will be plentiful incentives to resale or distribute this information \cite{hirshleifer1971private}. 
To capture the interplay between these features, here we introduce a simple model to study the spatial dynamics of aging information. 

%
%

\begin{figure}
\centering
\includegraphics[width = 1\columnwidth]{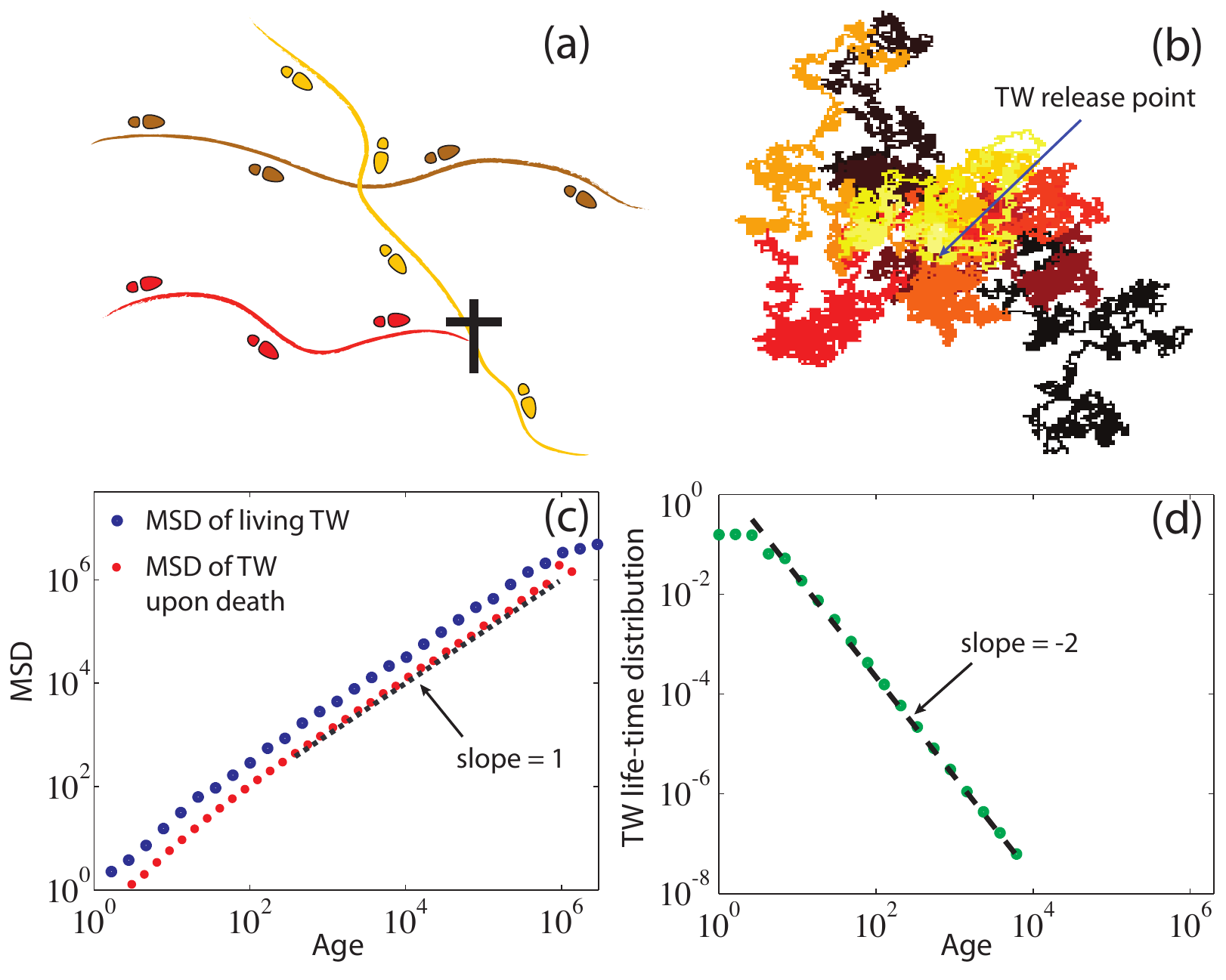}
\caption{Dynamics of time walkers. (a) Illustration of rules R1 and R2 applied to three walkers released at different times: old (brown), intermediate
(red), and young (yellow). The yellow TW can cross the traces of the brown and red TWs, but the red TW cannot cross the trace of the yellow one and is eliminated from the model.
(b) Snapshot of a simulation of TWs moving on a two-dimensional lattice. 
Lighter colors to indicate younger walkers.
(c) Mean square distance (MSD) of TWs as a function of their
age: Blue circles indicate distances reached by living TWs at a given age,
and red dots indicate distances reached and age obtained when eliminated (fitted by
MSD$\sim$age for old TWs).
(d) Distribution of TW lifetimes fitted by age$^{-2}$.
A java simulation of TW dynamics is found on-line at \cite{applet}. 
}
\label{tw}
\end{figure}

Researchers often use random walkers to model spreading of ideas, innovations, and rumors \cite{camagni1991innovation,rogers1995diffusion,valente1996network,brin1998anatomy,geroski2000models}. We also take the random walker as our starting point, but to capture the idea that the value of information decays with time and that new information in general is communicated with higher intensity than old, we extend the properties of the random walker and introduce the time walker (TW). In its simplest form, the TW is a random walker that is tagged with an age as a proxy for the value of the information it carries and that interacts with the trace left by other TWs in such a way that older walkers cannot cross the trace of younger walkers.

For simplicity and illustrative purposes, we emit our TWs from a single source on a two-dimensional lattice, but this framework can be generalized to accommodate multiple sources and to search with local information in both static and dynamic networks \cite{adamic2001search,rosvall2006self,rosvall2003modeling}. To model the interactions between TWs, we store two pieces of information at each lattice point: $(i)$ the age of the youngest TW to visit that point, and $(ii)$ an arrow pointing to the lattice point from which the youngest TW arrived. Below, we structure these data points as an information landscape and quantify their values by measuring how accurately a message can be routed back to the information source. We start, however, by detailing how the TWs move and update the information landscape:

R1. At every time step $t_i$ ($i=1,2,3,\ldots$), a new TW (TW$_i$) is born at the central point $(x,y)=(0,0)$ of the lattice and tagged with its birth time $t_i$, which it carries throughout its life.

R2. At every time step, all TWs make an attempt to step to a random neighboring lattice point and one of the events (a)-(c) occurs:
(a)
If TW$_j$ steps to a lattice point previously visited by TW$_k$, which stores newer information than TW$_j$ ($t-t_j > t-t_k$) carries, then TW$_j$ is eliminated from the model.
(b)
If TW$_k$ steps to a lattice point previously visited by TW$_j$, which stores older information than TW$_k$ ($t-t_k < t-t_j$) carries, then TW$_k$ completes the step and updates the age and the arrow at the lattice point (moving the arrow to point 
toward TW$_k$'s previous position).
(c)
If a TW visits the same lattice point more than once, the TW completes the step but does not update the arrow. We included this rule to ensure that all visited points connect back to the information source without closed loops. 

Figure \ref{tw} shows the dynamics of the TWs as they build and continuously update the information landscape (see also \cite{applet} for a java simulation). In Fig.~\ref{tw}(a), we illustrate how three TWs follow rules R1 and R2. The youngest TW (yellow) can cross the traces of the brown and the red TWs, but the red TW, which is older than the yellow TW, dies as soon as it encounters the trace of the yellow TW. Figure \ref{tw}(b) shows a snapshot of TW traces after a few hundred time steps. Hundreds of TWs have been born in the course of the simulation, but only about 20 are still present. We find that the spatial distribution of the TWs scales as that of normal random walkers. Figure~\ref{tw}(c) shows the mean square displacement of the time walkers at a given time (blue) and when they die (red). The constant difference between the two curves can be explained by the selection pressure on the TWs.
Those that survive walk away quickly from the centre, which is dominated by young walkers. The simulations show that those who prevail move only a fraction faster than the ones approaching from behind. This means that the older TWs do not need to move ballistically or super-diffusively in order to escape, which holds true also in higher dimensions. In Fig.~\ref{tw}(d), we show a histogram of TW lifetimes with asymptotic power-law decay, $p(t) \propto 1/t^2$. From panels (c) and (d), we conclude that most TWs die young and close to the source.

%
%
\begin{figure}
\centering
\includegraphics[width = \columnwidth]{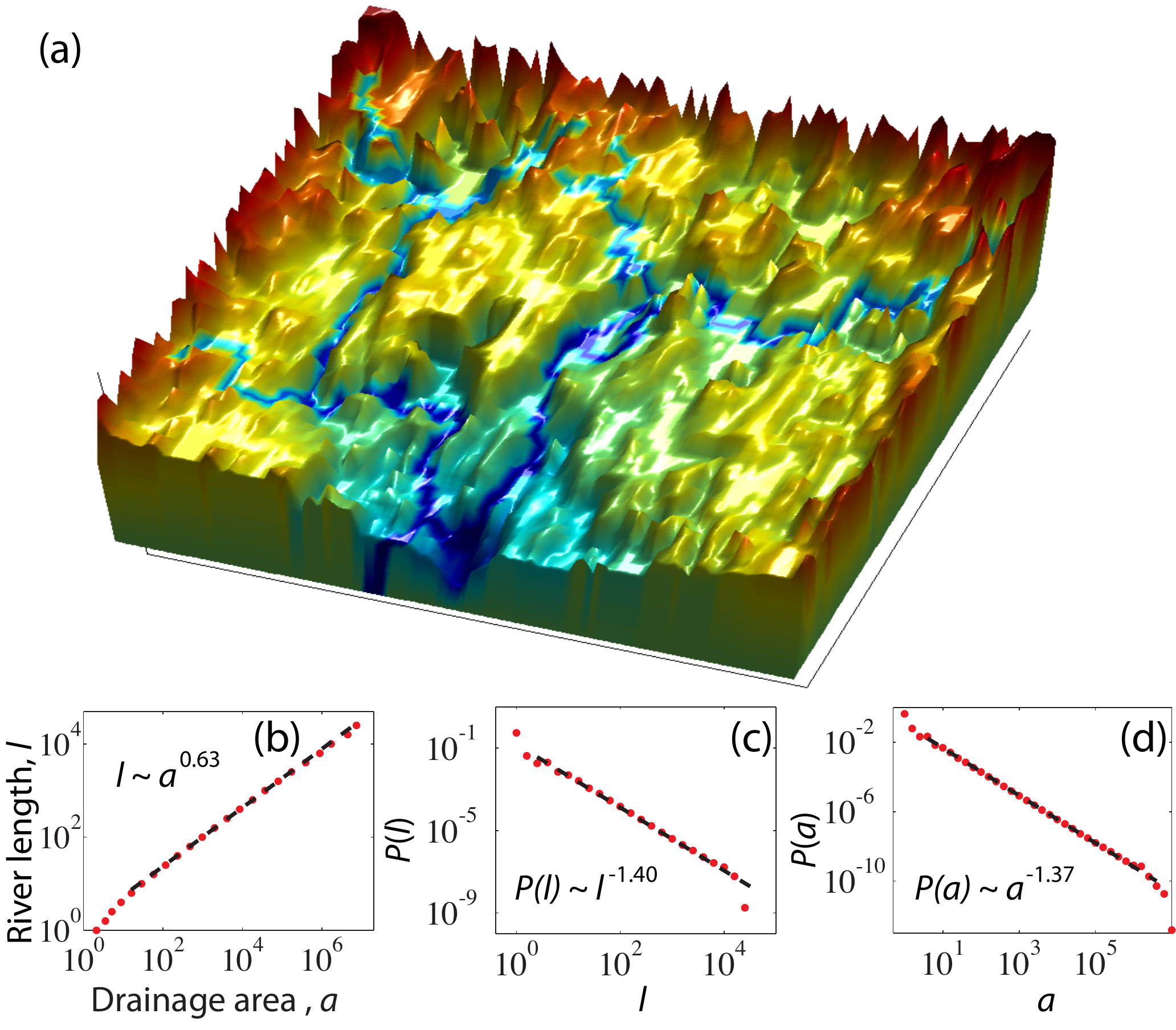}
\caption{
(a) Information landscape created by continually releasing TWs from a fixed point source (blue sink in foreground). Hills and crevasses represent old and young information. The age of each lattice point is equal to the age of the last TW that visited. Blue streaks show a few traces of tracking walkers, released close to the edge of the lattice, that follow the arrows pointing toward the source in the landscape, creating an information river network. Panels (b)-(d) contain statistics related to the information river network: (b) is the relation between upstream drainage area $a$ and the length $l$ of the longest upstream river (Hack's law), sampled over all lattice points. Panels (c) and (d) are histograms of $a$ and $l$, showing that they are scale-free. Dashed lines are linear fits. The development of the age landscape and the information river network are visualized in a java applet at \cite{applet}.
}
\label{ages}
\end{figure}

Until now we have focused on the spreading of information by means of the time walkers, but we now turn to the value of the information they left behind. As the TWs move over the lattice, they mark each lattice point with updated ages and arrows providing directions toward the TWs' source, which together effectively define the information landscape. Here we assume only that new information is more valuable than old, and rather than focusing on the value of the source information, we are interested in the information value of the landscape as a resource for providing efficient navigation back to the source to access up-to-date information. For example, we are interested in the value of a trader's extended network as a means of accessing private information, rather than how much money the trader can make with the private information itself. Figure \ref{ages}(a) illustrates the age profile of the information landscape as a surface plot with valleys, gradients and rivers. In general, since fewer steps are required to reach the central points than the remote corners of the lattice, points close to the information source have younger information than points farther way. However, the landscape can be highly intermittent, with plateaus of nearly constant age demonstrating the near space-filling property of random walkers in two dimensions, separated by sharp edges created by new TWs entering a region left untouched for a long time, leading to a huge age variance.

To quantify the value of the information left by the TWs, as measured by its age, we release passive tracking walkers on the lattice and study the paths they take from their release points back to the source point of the TWs, as they follow the arrows at each lattice point (see the blue rivers in Fig.~\ref{ages}(a)). By releasing a rain of trackers over the whole lattice, the myriad streams that correspond to the trajectories of the various trackers form an \emph{information river network}. We define a stream's depth at a given coordinate $(x,y)$ as the number of trackers that pass through $(x,y)$ if released upstream of $(x,y)$. In this way, the stream's depth is analogous to the upstream drainage area of a river of flowing water \cite{Hack,DORO}.

By evaluating the trajectories from a large number of tracking walkers, we can use the well-known scaling relations of actual rivers to characterize our information river network.  Figure~\ref{ages}(b) shows the relationship between the length $l$ of the longest river stream (equivalent to the largest number of tracking steps) within a drainage area and the size of the drainage area $a$. This is referred to as Hack's law \cite{Hack} in the literature and is given by $l\sim a^{h}$. For river networks of information, we find $h=0.63$, while in river networks of water (W), $h_{\rm W} \approx 0.57-0.60$ \cite{DORO}. Furthermore, both histograms of river lengths and drainage areas exhibit power-law distributions, $p(l)\sim l^{-\gamma}$ and $p(a)\sim a^{-\tau}$ with $\gamma=1.40$ and  $\tau = 1.37$ (see Fig.~\ref{ages} (c) and (d)).
Corresponding exponents for water river networks are $\gamma_{\rm W}=1.8\pm 0.1$ and $\tau_{\rm W} = 1.43\pm 0.02$ \cite{DORO} which means that the information rivers are in general longer, of more serpentine shape, and tend to have larger and wider drainage areas than real river networks.  
However, the scaling exponents associated with the information river
networks are only slightly smaller than $h_{\rm S} = 2/3$, $\gamma_{\rm
S} = 3/2$ and $\tau_{\rm S} = 4/3$ \cite{takayasu1988power}, which
characterize the directed rivers found in the Scheidegger model (S)
\cite{scheidegger1967stochastic}.

%
%

\begin{figure}
\centering
\includegraphics[width = 0.75\columnwidth]{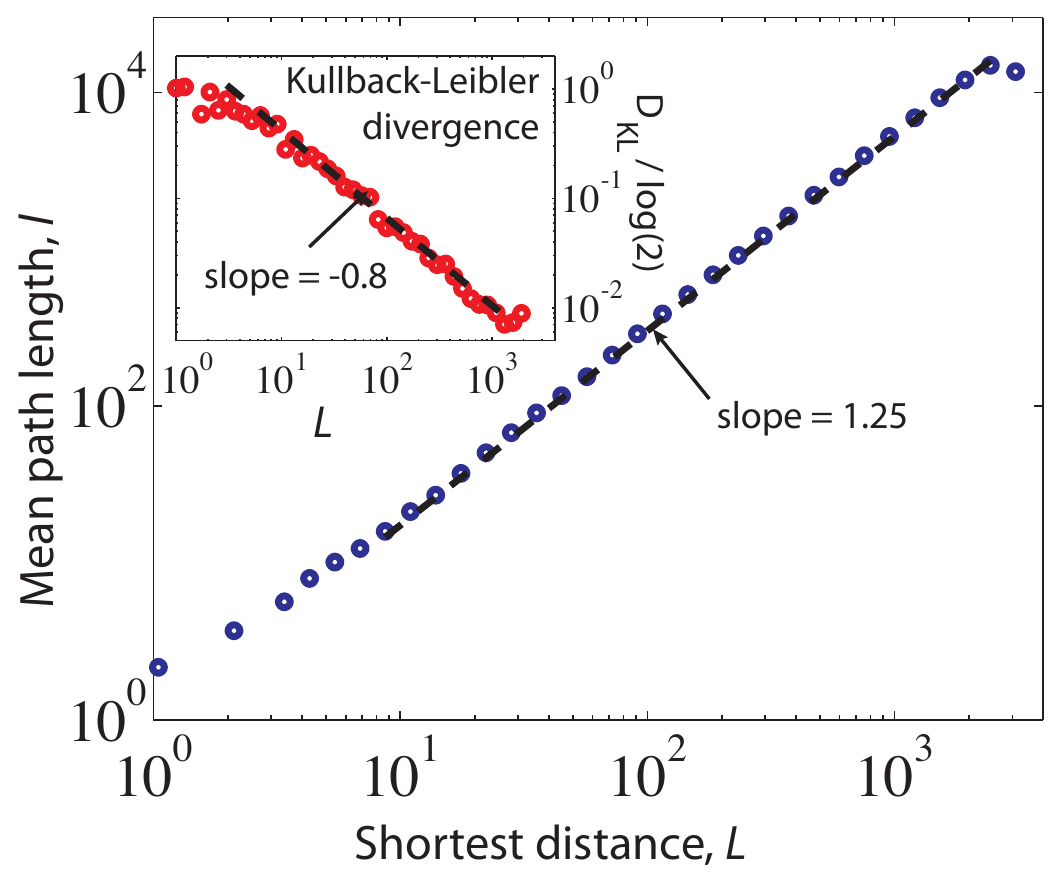}
\caption{Average path length of tracking walkers that follow the arrows in the information landscape back to the source as a function of the shortest distance to the source. The inset shows the corredponding decay in the Kullback-Leibler divergence $D_{\rm KL}$.}
\label{fig:MPL}
\end{figure}

The tracking walkers that are directed by the arrows in the landscape will, by construction, always find the information source independently of where they are released. This also holds true for an unbiased random search (in a finite system), but is much less efficient and scales unfavorably with system size. In Fig.~\ref{fig:MPL}, we show the mean search time for tracking walkers, quantified by their mean path length $l$ as function of the shortest (Euclidian) distance $L$ between the tracking walkers's starting point and the information source. The number of required steps scales asymptotically as $l\sim L^{1.25}$, which is much shorter than a random search~(RS) 
$l_{\rm RS} \sim \frac{1}{2} N \ln L$ \cite{SR} where $N$ is the number of points in the two-dimensional lattice.

Interestingly, the scaling exponent 1.25 for the tracking walkers matches the growth exponent for loop-erased random walks \cite{GL}, which has been shown analytically to be 5/4 in two dimensions \cite{RK}. The loop-erased random walk is a simple model for a self-avoiding walk derived from a random walk on a lattice with the rule that whenever the walker crosses its own trail, it removes the generated loop before it continues (very much like~R2). To our knowledge, the first proof of the 5/4 result in the physics literature appeared in ref.~\cite{majumdar1992lerw}, which established an exact mapping between the loop-erased random walk and the $q$-state Potts model.
After identifying the dynamic similarities between our search process and the loop-erased random walk, we can extend our results to higher dimensions. For example, in three dimensions $l\sim L^{1.62}$ \cite{guttmann1990critical} with search lengths that are far superior to a random search from which there is a finite chance that an explorer will never return.

Even though the information left by the TWs dramatically reduces the search time for the tracking walkers, their walks remain far from ballistic, with a scaling exponent well above 1.0. The reason is that the paths followed by the tracking walkers become increasingly erratic the farther away the walkers are from the source, and random meanderings make the mean path length substantially longer. We quantify the inaccuracy of the arrows with the Kullback-Leibler divergence $D_{KL}$ \cite{kullback}, which measures the information gain in using the directional information left by the TWs. With $p_{\rm TW}$ for the fraction of arrows (calculated from an ensemble of landscapes) pointing to a lattice point located closer to the information source, we have
$D_{KL} = p_{\rm TW} \log (p_{\rm TW}/q) - 
(1-p_{\rm TW}) \log \left((1-p_{\rm TW})/q\right)$, 
where $q$ is the likelihood the tracking walker will step in the correct direction (that is, closer to the source) by chance, i.e.~$q=0.5$ on all lattice points except the cross through the centre, for which $q=0.25$. The inset in Fig.~\ref{fig:MPL} shows the decay in $D_{KL}$ as a function of $L$, and the linear fit indicates that $p_{\rm TW} \sim q+{\rm const.}/L^{0.8}$.

%
%

The declining precision of the arrows with distance from the target is associated with the decreasing density of TWs as they get farther from the source; the on-going selection of TWs is essential for maintaining order in the spreading of information. Consequently, a high density of TWs or a strong selection pressure generates an efficient search landscape. To explore this effect, and to investigate the robustness of our model, we introduce replicating TWs, TWs that can reproduce and have a chance to die when encountering the trace of an older walker (but still die with certainty when crossing a younger TW's trail). Conforming to our interpretation of the TWs as carriers of information, a replicating TW corresponds to information broadcasting while TWs dying from an old trail crossing correspond to communication failures.

\begin{figure}
\centering
\includegraphics[width=1.0\columnwidth]{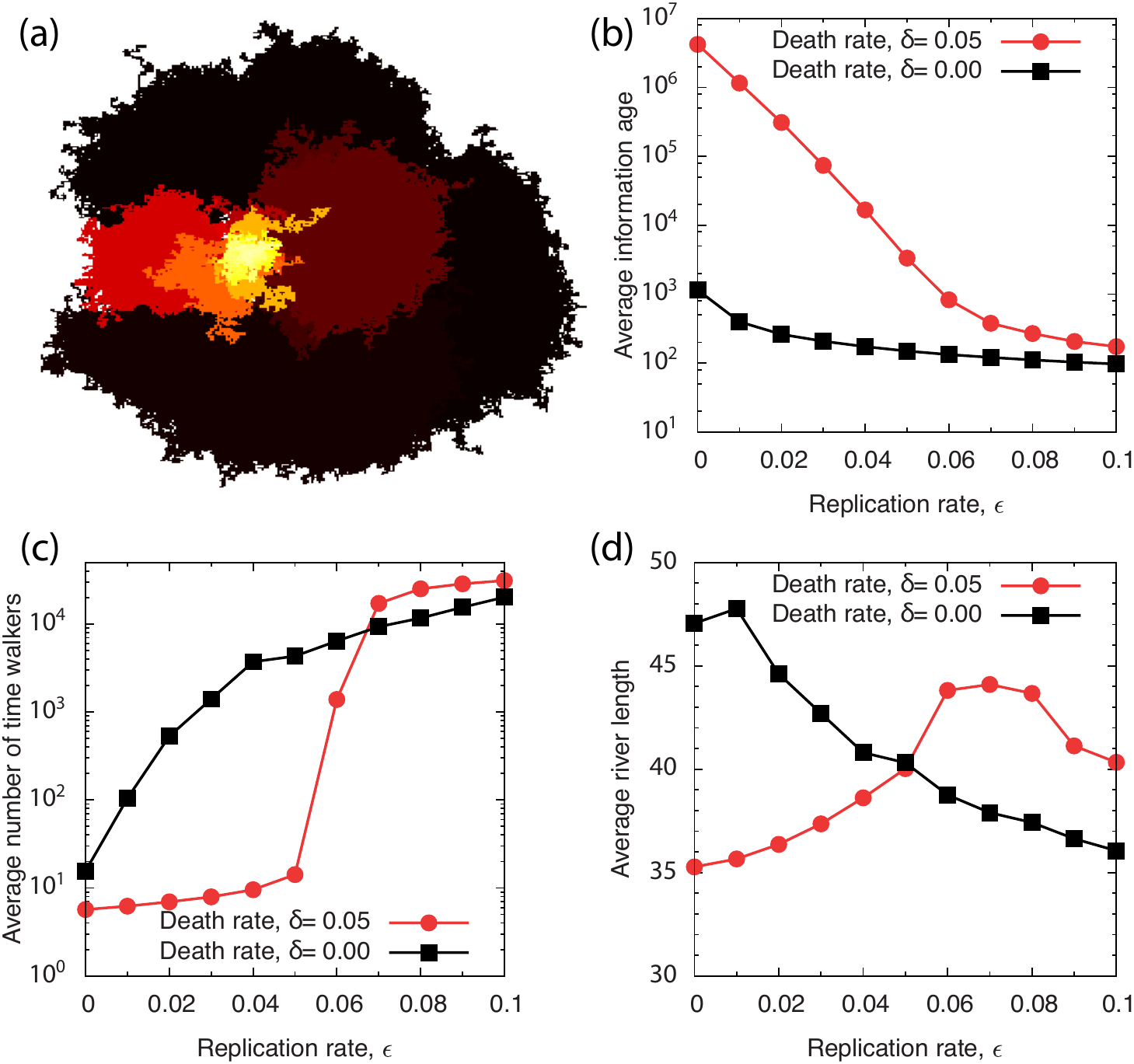}
\caption{Dynamics of TWs that at each step replicate with probability $\epsilon$ and die with probability $\delta$.
(a) Snapshot of a simulation on a two-dimensional lattice, with lighter colors indicating younger TWs ($\epsilon=0.01$ and $\delta=0.02$).
(b) Average information age of the entire lattice decreases with increasing
replication rate, regardless of death rate $\delta\geq 0$.
(c) The average number of TWs increases rapidly when the replication rate 
exceeds the death rate $\epsilon > \delta$.
(d) The average information river length reaches a maximum when the replication
rate is slightly larger than the death rate.
}
\label{branchdeath}
\end{figure}

In Fig.~\ref{branchdeath}, we show that replicating TWs with the replication rate $\epsilon$ and death rate $\delta$ (when meeting an old trail) indeed can generate more efficient information landscapes. Figure~\ref{branchdeath}(a) shows how wave fronts of replicating TWs of the same age efficiently fill the space behind older ones as they move out from the source. Considering replication only ($\delta = 0$), the average information age and the average river length decrease with an increasing replication rate, but at the cost of a hugely increased number of active TWs (see Fig.~\ref{branchdeath}(b)-(d)). Spontaneously dying TWs ($\delta > 0$) reduce this communication cost (see Fig. \ref{branchdeath}(c)). However, when the death rate becomes larger than the replication rate, the TWs cannot reach distant parts of the lattice (they die off exponentially), which results in a very large average information age (see Fig.~\ref{branchdeath}(b)). For $\epsilon = 0$, typical paths are short but the coverage of the lattice is small. We conclude that the relatively directed paths obtained by spontaneously dying walkers comes at the cost of a system that would be very slow to adapt to situations in which, for example, the source point moves or the lattice becomes damaged.

%
%

In summary, in this letter we have introduced the time walker (TW), derived from the random walker, as a method of studying the spatial dynamics of aging information. The movements of interacting TWs capture the concept that new information is often more valuable than old information. The time walkers interact with the trace left behind by other walkers, and the full system of traces of different ages forms a navigable, self-similar, information landscape. We have visualized searches back to the information source in this landscape as an information river network, and showed that the search lengths scales as the length of loop-erased random walks. We have also studied the effects of broadcasting and communication failures with replicating and dying TWs, and showed that they can generate more efficient information landscapes, but at a much higher communication cost.

In the two-dimensional system with large distances that we have studied here, the communication costs of generating efficient information landscapes quickly diverge with system size. But distances are much shorter in social networks, even for large systems \cite{milgram}, and the very simple communication mechanism captured by a TW can generate efficient information landscapes. This suggests an alternative way in which to interpret Stanley Milgram's famous six degrees of separation. 

\begin{acknowledgements}
This work was supported by the Danish National Research Foundation through the Center for Models of Life.
\end{acknowledgements}

\end{document}